\documentclass[runningheads]{llncs} % 

\usepackage[T1]{fontenc} % 
\usepackage{graphicx}
\usepackage{amsmath}
\usepackage{bm}
\usepackage{nccmath}
\usepackage{float}
\begin{document}

% -------------------------------------------------------------------------
% Header: Title, Authors, Institute [cite: 29-87]
% -------------------------------------------------------------------------
\title{Learning about Corner Kicks in Soccer by Analysis of Event Times Using a Frailty Model}

% Use \titlerunning{...} if the title is too long for the header [cite: 44]
\titlerunning{Frailty Model for Corner Kicks}

% Authors separated by \and 
% Use \inst{1} if authors have different affiliations [cite: 55]
\author{Riley L. Isaacs \and X. Joan Hu \and K. Ken Peng \and Tim B. Swartz}

% Short author list for the header [cite: 70]
\authorrunning{R. Isaacs et al.}

% Affiliations are required in LLNCS. 
% Please replace <Institute> with your actual affiliation.
\institute{Statistics and Actuarial Science/Simon Fraser University}

\maketitle % [cite: 90]

% -------------------------------------------------------------------------
% Abstract and Keywords [cite: 93-98]
% -------------------------------------------------------------------------
\begin{abstract}
Corner kicks are an important event in soccer because they are often the result of strong attacking play and can be of keen interest to sports fans and bettors. Peng, Hu, and Swartz (2024, {\em Computational Statistics}) formulate the mixture feature of corner kick times caused by previous corner kicks, frame the commonly available corner kick data as right-censored event times, and explore patterns of corner kicks. This paper extends their modeling to accommodate the potential correlations between corner kicks by the same teams within the same games. We consider a frailty model for event times and apply the Monte Carlo Expectation Maximization (MCEM) algorithm to obtain the maximum likelihood estimates for the model parameters. We compare the proposed model with the model in Peng, Hu, and Swartz (2024) using likelihood ratio tests. The 2019 Chinese Super League (CSL) data are employed throughout the paper for motivation and illustration.

% Keywords must be separated by \and [cite: 97]
\end{abstract}

% \newpage % LLNCS usually flows continuously; commented out but can be kept if necessary.

\section{Introduction}

Soccer corner kicks are awarded to the attacking team after the defending team puts the ball into touch across their own goal line. In elite soccer a very small percentage of corner kicks lead to goals, Casal et al (2015) Yiannakos and Armatas (2006) found that more goals tend to be scored as a result of corner kicks than any other type of set piece. The stoppage in play allows both teams the opportunity to strategize, with the attacking team making decisions such as what depth and height to kick the ball, while the defending team decides which players or areas of the field to defend.\\

Corner kicks have a breadth of literature, including Casal et al (2015) which investigates the efficacy of different types of corner kicks and investigates related variables. Liu et al (2015) includes corner kicks as one of the performance variables when building team strength profiles. Maneiro et al. (2021) analyses the change in corner kick strategies over three consecutive world cups.\\

This paper is a direct extension of Peng, Hu, Swartz (2024) which investigates the time elapsed since a starting point until a corner kick using event history analysis. Event history analysis is most commonly used in biomedical research, which faces many of the same challenges that are presented in our analyses of corner kicks. Peng, Hu and Swartz (2024) addressed concerns regarding the right censoring of corner kicks. For a comprehensive summary on the methods commonly used to address right censoring in event time analysis, see Kalbfleisch and Prentice (2011). In the context of corner kicks right censoring occurs when a goal is scored, or the end of a half or game occurs. This stops play before the corner kick can be observed.\\

Peng, Hu and Swartz (2024) also addressed the fact that there is a a non-ignorable probability of a corner kick almost immediately following another corner kick, typically as a result of a defensive deflection which goes out of bounds. This phenomenon means conventional time to event distributions are not reasonably suited to our corner kick problem. As in their initial paper we employ a mixture distribution to address the issue. McLachlan et al. (2019) reviews the theory and methodological developments of finite mixture models.\\

This paper extends the model of Peng, Hu and Swartz (2024) by exploring two different frailty models. These models are more general as the do not require the assumption that time to corner kicks are independent of each other. Frailty models are commonly used to model correlated event times in the life sciences, with Vaupel et al. (1979) introducing the notion of frailty to account for heterogeneity between groups when modelling mortality. It has also been used in reliability engineering as in Ghomghaleh et al (2020) where a frailty model is shown to outperform the classical model for determining the remaining useful life of an excavator.\\

The 2019 Chinese Super League data includes the event history data for 233 games, which includes 2314 corner kicks. For the purpose of this analysis the corner kicks are split into Type 1 kicks which are corner kicks that are preceded by a starting event that is not a corner kick, and Type 2 kicks which are preceded by another corner kick.\\

Section 2 begins by introducing the structure of the data, and distinguishes the latent variables from those which are observed. Then the frailty model is introduced. Section 3 covers the special case that occurs when the latent frailty variables are gamma distributed and provides an MCEM algorithm that can be used to obtain maximum likelihood estimates. Section 4 describes our analysis of the 2019 Chinese Super League (CSL) data and a simulation study before we wrap up by discussing other methods for analyzing time to corner kick data.\\

\section{Statistical Methodology}
\subsection{Data Formulation}
For a set of event history data containing $I$ teams we use $i$ to denote the $ith$ team, each of the $I$ teams play $J_i$ games with $ij$ denoting the $ith$ team's $jth$ game. In the $ith$ team's $jth$ game there are $n_{ij}$ events where they are either awarded a corner kick or a censoring event occurs such as a goal, or end of half. Each of the events where either a corner kick or censoring event occurs is assigned an index $k$. We let the random variables $T_{ijk}$ and $C_{ijk}$ represent the time to the corner kick from previous event and time to censoring event from the previous event, respectively. However we only observe $Y_{ijk}=min(T_{ijk},C_{ijk})$ Additionally we assume non-informative censoring time. We also define an indicator of whether the corner kick is observed $\delta_{ijk}=I(Y_{ijk}=T_{ijk})$. Finally for each event we define $Z_{ijk}$ to be a vector of covariates that may include whether or not the team is at home, and score differential at the time of the corner kick. 
We define 
$OD = \{y_{ijk},z_{ijk},\delta_{ijk}:i=1,...,I, j=1,...,J_i, k=1,...,n_{ij}\}$.
to be the set of observed data.\\

As in Peng, Hu, Swartz (2024) we define Type 1 corner kicks to be a corner kick that occurs after play is started by any event that is not a corner kick. Type 2 corner kicks are corner kicks that occur after play is restarted by a corner kick. Type 2 corner kicks are broken down further into two categories, the first being type two short (Type 2-S) which is a corner kick that was generated directly by the preceding corner kick. There are also type two long corner kicks (Type 2-L) which are not generated directly by the previous corner kick and tend to have longer gap times. For every $Y_{ijk}$ such that $\delta_{ijk}=0$ we define the latent variable $\eta=I(Y_{ijk}| CK\, is\, Type\, 2-L)$. This would be observable with tracking data, or even potentially with very detailed event history data. However, for the purpose of this paper we consider it to be unobservable. We define our first "full" set of data to be 
$FD1 = \{y_{ijk},z_{ijk},\delta_{ijk},\eta_{ijk}:i=1,...,I, j=1,...,J_i, k=1,...,n_{ij}\}$. This is the data considered in Peng, Hu, Swartz (2024).\\

Let $W_{ij}$ be a latent frailty random variable associated with the $ith$ team's $jth$ game. We define our second set of full data 
$FD2 = \{y_{ijk},z_{ijk},\delta_{ijk},\eta_{ijk},w_{ij}:i=1,...,I, j=1,...,J_i, k=1,...,n_{ij}\}$.

\subsection{Statistical Model}
In (Peng, Hu, Swartz 2024) they specify two unique parametric proportional hazards models as follows. For Type 1 and Type 2-L kicks the hazard is
\begin{equation}
    h_1^{(0)}(t|\bm{z};\gamma_1,\lambda_1,\bm{\beta})= \gamma_1\lambda_1(\lambda_1t)^{(\gamma_1-1)}e^{\bm{z'\beta}}.
\end{equation}
The hazard function used for Type 2-S corner kicks is 
\begin{equation}
    h_2^{(0)}(t|\gamma_2,\lambda_2)= \gamma_2\lambda_2(\lambda_2t)^{(\gamma_2-1)}.
\end{equation} 
This model is suited to data where all corner kicks are independent of one another. Analysis using this model requires that we assume independence of corner kicks. In reality, it is likely that a team's corner kicks within the same game are correlated due to their strategy, personnel, and form on a given day. Using a model that requires weaker assumptions about the correlation structure of the data would allow for valid inference provided the corner kicks are not all independent of each other. To avoid making the assumption that all corner kicks are independent we introduce the frailty random variable $W_{ij}$ which acts multiplicatively on the hazard function for the $ith$ team's Type 1 and Type 2-L corner kicks in their $jth$. Resulting in the following hazard functions:
\begin{equation}
    h_1(t_{ijk}|\bm{z}_{ijk},w_{ij};\gamma_1,\lambda_1,\bm{\beta})= w_{ij}\gamma_1\lambda_1(\lambda_1t_{ijk})^{(\gamma_1-1)}e^{\bm{z_{ijk}'\beta}},
\end{equation}
\begin{equation}
and \ 
    h_2(t_{ijk}|\gamma_2,\lambda_2)= \gamma_2\lambda_2(\lambda_2t_{ijk})^{(\gamma_2-1)}.
\end{equation} 
Let $S_1(t|\bm{z},w;\gamma_1,\lambda_1,\bm{\beta})$ and $S_2(t|\gamma_2,\lambda_2)$ be the unique survival functions determined by $h_1(t|\bm{z},w;\gamma_1,\lambda_1,\bm{\beta})$ and $h_2(t|\gamma_2,\lambda_2)$ respectively.\\

Under this formulation if $w_{ij} \equiv 1$, our model $h_1(t_{ijk}|\bm{z}_{ijk},w_{ij};\gamma_1,\lambda_1,\bm{\beta})$ reduces to $h_1^{(0)} (t_{ijk}|\bm{z}_{ijk};\gamma_1,\lambda_1,\bm{\beta})$. The proposed model may well formulate data where the timing of corner kicks taken by the same team within the same game are more similar to each other than other corner kicks. Meaning that if factors such as strategy, personnel and match day form do impact the timing of corner kicks this model will be more suitable than previous models.

\subsection{Likelihood}
For the kth observed time in the $ith$ team's $jth$ game, assume
\begin{equation}
    P(\eta_{ijk}=1|z_{ijk};\alpha_0,\bm{\alpha})= \pi(z_{ijk};\alpha_0,\bm{\alpha})= \frac{1}{1+e^{-(\alpha_0+\bm{z_{ijk}'\alpha})}}.
\end{equation}
We assume that all corner kick times not taken by the same team within the same game are independent, that the $W_{ij}$ are independently sampled from a distribution with pdf $f_W(.)$, and that given $W_{ij}$ all of the $ith$ team's corner kicks in their $jth$ game are independent. In the case of univariate frailty models the frailty distribution is often specified to be the gamma distribution, positive stable distribution, compound poisson distribution, or the log-normal distribution (Weike, 2003). Section two of Balan and Putter (2020) discusses some effects of the choice in frailty distribution.\\ 

The likelihood contribution of the $ith$ team's $jth$ game given data $FD2$ is
\begin{multline}
L^{(1)}_{ij}(\bm{\theta}) =
f_W(w_{ij})
\Pi^{n_{ij}}_{k=1}
L^{(1)}_{ijk}(\bm{\theta})
[\pi(z_{ijk};\alpha_0,\alpha)^{\eta_{ijk}}
(1-\pi(z_{ijk};\alpha_0,\alpha))^{1-\eta_{ijk}}]^{\delta_{ij,k-1}},
\end{multline}
where 
\begin{multline*}
L^{(1)}_{ijk}(\bm{\theta}) = [(h_1(y_{ijk}|\mathbf{z_{ijk}},w_{ij};\bm{\theta}_1))^{\delta_{ijk}}
S_1(y_{ijk}|\mathbf{z_{ijk}},w_{ij};\bm{\theta}_1)]^{1-\delta_{ij,k-1}}\\
[((h_1(y_{ijk}|\mathbf{z_{ijk}},w_{ij};\bm{\theta}_1)^{\eta_{ijk}} h_2(y_{ijk}|\bm{\theta}_2)^{(1-\eta_{ijk})})^{\delta_{ijk}} \\
(S_1(y_{ijk}|\mathbf{z_{ijk}},w_{ij};\bm{\theta}_1)^{\eta_{ijk}} S_2(y_{ijk}|\bm{\theta}_2)^{(1-\eta_{ijk})}))]^{\delta_{ij,k-1}}.
\end{multline*}
The likelihood considering all games given data $FD2$ is
\begin{equation}
    L^{(1)}(\bm{\theta}) = \Pi^I_{i=1}\Pi^{J_i}_{j=1}L^{(1)}_{ij}(\bm{\theta}).
\end{equation}
In reality we do not observe $FD2$, and we only observe data $OD$. In general there is no marginal closed form of the likelihood function conditional only on $OD$. However, when all $W_{ij}$ are independently sampled from a gamma distribution, there exists a closed form for the marginal likelihood given data $FD1$, this special case is discussed in detail in Section 4. 

\section{Computing Maximum Likelihood Estimates}
$L^{(1)}(\bm{\theta})$ is conditional on $FD2$ which contains latent data making it impossible to analytically compute maximum likelihood estimates. Neath (2012) demonstrates that the Monte Carlo Expectation Maximization algorithm converges to the maximum of the observed data likelihood. This property makes MCEM a suitable choice for numerically computing the MLE.
~\\
\hrule
MCEM Algorithm for General Frailty Distribution
\hrule
Given the estimate at $d$ iterations $\theta^{(d)}$, with $\theta^{(0)}$ for the initial value and an initial sample of $\eta$s $\eta_{ij1}^{(0)},...,\eta_{ijn_{ij}}^{(0)}$.\\
1: \textbf{Repeat}\\
1.1 MCE-step. Let $l^{(1)}(\theta)=log(L^{(1)}(\theta))$, approximate $Q(\theta,\theta^{(d)})=E(l^{(1)}(\theta)|FD2; \theta^{(d)})$
using the sample mean $\Tilde{Q}(\theta,\theta^{(d)}) = \frac{1}{M}\Sigma_{m=1}^M l^{(1)}(\theta)$ conditional on $\{(y_{ijk},\delta_{ijk},\bm{z_{ijk}},\eta_{ijk}^{(m)},w_{ij}^{(m)}:k=0,...,n_{ij},j=1,...,J_i,i=1,...,I\}$.
For $m=1,...,M$\\
a) Sample $w_{ij}^{(m)^{(d+1)}}$ from $f_W(w|\eta_{ij1}^{(m)^{(d)}} ,...,\eta_{ijn_{ij}}^{(m)^{(d)}},OD;\bm{\theta}^{(d)})$\\
b) Sample $\eta_{ijk}^{(m)^{(d+1)}}$ from a $Bernoulli(P)$ distribution, where P is the conditional probability of a corner kick being Type 2-L given the observed time of the corner kick. $P=\frac{P_1}{P_1+P_2}$, where\\
\begin{equation*}
    P_1 = \pi(\mathbf{z}_{ijk};\alpha_0^{(d)},\bm{\alpha}^{(d)})
(h_1(y_{ijk}|w_{ij}^{(m)^{(d+1)}},\mathbf{z}_{ijk};\bm{\theta}^{(d)}))^{\delta_{ijk}}S_1(y_{ijk}|w_{ij}^{(m)^{(d+1)}},\mathbf{z}_{ijk};\bm{\theta}^{(d)}),
\end{equation*}
\begin{equation*}
   and \ P_2 = \pi(\mathbf{z}_{ijk};\alpha_0^{(d)},\bm{\alpha}^{(d)})
(h_1(y_{ijk}|\bm{\theta}^{(d)}))^{\delta_{ijk}}S_1(y_{ijk}|\bm{\theta}^{(d)}).
\end{equation*}
Note $\eta_{ijk}$ is not defined for Type 1 corner kicks ($\delta_{ij,k-1}$=0), in these instances there is no need to sample $\eta_{ijk}$.\\
1.2 M-step. Update the estimate of $\theta$ as $\theta^{(d+1)}=argmax_{\theta}\Tilde{Q}(\theta,\theta^{(d)})$ \\
2. \textbf{Until} $\{\theta^{(d)},d=1,2,...\}$ converges. We consider the algorithm to converge when the change of 2-norm of $\theta^{(d+1)}-\theta^{(d)}$ is smaller than some tolerance.
\hrule
\subsection{Standard Errors}
Luois (1982) provides a useful expression for the Hessian of the log posterior distribution for the EM algorithm. Using the data from the last step of the MCEM algorithm we numerically compute the following hessian matrix:
\begin{equation*}
    \frac{1}{M}\Sigma_{m=1}^{M}\frac{D^2log(L^{(1)}(\bm{\theta}))}{D^2\bm{\theta}} + 
    \frac{1}{M}\Sigma_{m=1}^{M}(\frac{Dlog(L^{(1)}(\bm{\theta}))}{D\bm{\theta}})^2 -
    [\frac{1}{M}\Sigma_{m=1}^{M}\frac{Dlog(L^{(1)}(\bm{\theta}))}{D\bm{\theta}}]^2
\end{equation*}
The inverse of this matrix is the covariance matrix of which the diagonal entries provide standard errors for our parameter estimates.
\subsection{Example of Frailty Distribution: Log-Normal}
The log-normal distribution is a commonly used frailty distribution, it is not a member of the power variance family (PVF) which contains the gamma distribution and other frequently used frailty distributions (Balan and Putter, 2020). Constraining the mean to be one gives the following one-parameter Log-Normal distribution:
\begin{equation*}
    f_W(w) = \frac{exp(\frac{-(ln(w)+\frac{\sigma_w^2}{2})^2}{2\sigma_w^2})}{w\sigma_w\sqrt{2\pi}}.
\end{equation*}
We develop the following procedure for sampling from $f_W(w|FD1;\bm{\theta})$ which can be applied to part a) of the MCEM algorithm described earlier in this section.\\
Using the fact that $f_W(w_{ij}|FD1;\bm{\theta})\propto L^{(1)}_{ij}(\theta),$
\begin{equation*}
    f_W(w_{ij}|FD1;\bm{\theta}) \propto r(w_{ij}) = exp(\frac{-(ln(w_{ij})+\frac{\sigma_w^2}{2})^2}{2\sigma_w^2})
    w_{ij}^{\phi-1}
    e^{-w_{ij}\psi}
\end{equation*}
where $\phi=\Sigma_{k=1}^{n_ij}[\delta_{ijk}(1-\delta_{ijk-1})+\eta_{ijk}\delta_{ijk}\delta_{ijk-1}]$ and 
$\psi=\Sigma_{k=1}^{n_ij}(\lambda_1y_{ijk})^{\gamma_1}
    (1-\delta_{ijk-1}+\eta_{ijk}\delta_{ijk})$.\\
We can use rejection sampling to sample frailty variables conditional on the data without needing to find the normalizing constant. The following is a rejection sampling protocol for this case.\\
~\\
\hrule
Rejection Sampling Algorithm
\hrule
1. If $\phi$ and $\psi$ are both greater than zero sample $u$ from a $Gamma(\phi,\psi)$ distribution and $v$ from a $Unif(0,1)$ distribution. Else, sample $u$ from a $Gamma(\phi+1,\psi+1)$ distribution and $v$ from a $Unif(0,1)$ distribution. Let $g(u)$ denote the pdf of the selected proxy distribution.\\
2. If $v < \frac{r(u)}{g(u)}$, let $w_{ij}=u$ and stop, else repeat.
Doing this M times obtains a sample of size M from the conditional distribution of $w_{ij}$.
\hrule

\section{Special Case: Gamma Frailty}
With the added assumption that all $W_{ij}$ are independently sampled from a $Gamma(\frac{1}{\theta_w},\frac{1}{\theta_w})$ distribution with pdf
\begin{equation}
    f_W(w) = \frac{(\frac{1}{\theta_w})^{\frac{1}{\theta_w}}}{\Gamma(\frac{1}{\theta_w})}w^{\frac{1}{\theta_w}-1}_{ij}e^{-\frac{w_{ij}}{\theta_w}},
\end{equation}
it is possible to analytically integrate the frailty random variables out and obtain a marginal likelihood distribution conditional on $FD1$.\\

This distribution was chosen for its useful mathematical properties. It's mean is $1$ meaning that the hazards for each team's game in our model vary around $h_1^{(0)}(t|\bm{z};\gamma_1,\lambda_1,\bm{\beta})$. The variance of this distribution is $\theta_w$ which is easily interpretable, it also means that as $\theta_w$ approaches zero, $f_W(w)$ approaches a point mass at $1$. Intuitively this means the closer $\theta_w$ is to $0$ the less hazard functions vary. Most importantly this distribution is selected because there exists a closed form for the marginal likelihood conditional on $FD1$.

The contribution of the $ith$ team's $jth$ game to the likelihood conditional on $FD1$ is
\begin{multline}
L^{(2)}_{ij}(\bm{\theta}) = \frac{(\frac{1}{\theta_w})^{1/\theta_w}}{\Gamma(1/\theta_w)}
\frac{\Gamma(\alpha^{(2)})}{(\beta^{(2)})^{\alpha^{(2)}}}
\Pi^{n_{ij}}_{k=1}L^{(2)}_{ijk}(\bm{\theta}) 
[\pi(z_{ijk};\alpha_0,\alpha)^{\eta_{ijk}}
(1-\pi(z_{ijk};\alpha_0,\alpha))^{1-\eta_{ijk}}]^{\delta_{ij,k-1}},
\end{multline}
where
\begin{multline*}
    L^{(2)}_{ijk}(\bm{\theta}) = [(h_1(y_{ijk}|\mathbf{z_{ijk}};\gamma_1,\lambda_1,\bm{\beta}))^{\delta_{ijk}}
S_1(y_{ijk}|\mathbf{z_{ijk}};\gamma_1,\lambda_1,\bm{\beta})]^{1-\delta_{ij,k-1}}\\
[((h_1(y_{ijk}|\mathbf{z_{ijk}};\gamma_1,\lambda_1,\bm{\beta})^{\eta_{ijk}} h_2(y_{ijk}|\gamma_2,\lambda_2)^{(1-\eta_{ijk})})^{\delta_{ijk}} \\
(S_1(y_{ijk}|\mathbf{z_{ijk}};\gamma_1,\lambda_1,\bm{\beta})^{\eta_{ijk}} S_2(y_{ijk}|\gamma_2,\lambda_2)^{(1-\eta_{ijk})}))]^{\delta_{ij,k-1}}
\end{multline*}
and
\begin{equation*}
    \alpha^{(2)}=\frac{1}{\theta_w}+\Sigma_{k=1}^{n_{ij}}\delta_{ij,k}(1-\delta_{ij,k-1}+\eta_{ijk}\delta_{ij,k-1})
\end{equation*}

\begin{equation*}
    \beta^{(2)}=\frac{1}{\theta_w}+\Sigma_{k=1}^{n_{ij}}(\lambda_1y_{ijk})^{\gamma_1}e^{\bm{z}_{ijk}\bm{\beta}}(1-\delta_{ij,k-1}+\eta_{ijk}\delta_{ij,k-1}).
\end{equation*}
That yields the following marginal likelihood:
\begin{equation}
    L^{(2)}(\theta) = \Pi^I_{i=1}\Pi^{J_i}_{j=1}L^{(2)}_{ij}(\theta).
\end{equation}
$L^{(2)}(\theta)$ is conditional on $FD1$ which still contains latent data making it impossible to obtain Maximum Likelihood Estimates by directly maximizing $L^{(2)}(\theta)$. The MLE can be obtained using $L^{(2)}(\theta)$ by implementing a slighty modified version of the MCEM algorithm outlined in section 2.3.

\subsection{Computing Maximum Likelihood Estimates (MLE)}
By integrating out the frailty random variables we obtained $L^{(2)}(\theta)$ which is conditional on $FD1$. This allows us to simplify the MCEM algorithm and remove a step, because there is no longer the need to simulate the frailty random variables. The following is the simplified MCEM algorithm.
~\\
\hrule
MCEM Algorithm for Gamma Distributed Frailty Random Variables
\hrule
Given the estimate at $d$ iterations $\theta^{(d)}$, with $\theta^{(0)}$ for the initial value.\\
1: \textbf{Repeat}\\
1.1 MCE-step. Let $l^{(2)}(\theta)=log(L^{(2)}(\theta))$, approximate $Q(\theta,\theta^{(d)})=E(l^{(2)}(\theta)|FD2; \theta^{(d)})$
using the sample mean $\Tilde{Q}(\theta,\theta^{(d)}) = \frac{1}{M}\Sigma_{m=1}^M l^{(1)}(\theta)$ conditional on $\{(y_{ijk},\delta_{ijk},\bm{z_{ijk}},\eta_{ijk}^{(m)}:k=0,...,n_{ij},j=1,...,J_i,i=1,...,I\}$.\\
For $m=1,...,M$\\
Sample $\eta_{ijk}$ from a $Bernoulli(P)$ distribution, where P is the conditional probability of a corner kick being Type 2-L given the observed time of the corner kick. $P=\frac{P_1}{P_1+P_2}$, where\\
\begin{equation*}
    P_1 = \pi(\mathbf{z}_{ijk};\alpha_0^{(d)},\bm{\alpha}^{(d)})
(h_1(y_{ijk}|w_{ij},\mathbf{z}_{ijk},\bm{\theta}^{(d)}))^{\delta_{ijk}}S_1(y_{ijk}|w_{ij},\mathbf{z}_{ijk},\bm{\theta}^{(d)})
\end{equation*}
\begin{equation*}
    and \ P_2 = \pi(\mathbf{z}_{ijk};\alpha_0^{(d)},\bm{\alpha}^{(d)})
(h_1(y_{ijk}|\bm{\theta}^{(d)}))^{\delta_{ijk}}S_1(y_{ijk}|\bm{\theta}^{(d)}).
\end{equation*}
Note $\eta_{ijk}$ is not defined for Type 1 corner kicks ($\delta_{ij,k-1}$=0), in these instances there is no need to sample $\eta_{ijk}$.\\
1.2 M-step. Update the estimate of $\theta$ as $\theta^{(d+1)}=argmax_{\theta}\Tilde{Q}(\theta,\theta^{(d)})$ \\
2. \textbf{Until} $\{\theta^{(d)},d=1,2,...\}$ converges. We consider the algorithm to converge when the change in the 2-norm of $\theta^{(d+1)}-\theta^{(d)}$ is less than a desired tolerance.
\hrule
~\\
We employed the nlminb function to compute the maximum of the estimated-expected log-likelihood at each iteration of the MCEM algorithm. While various r functions can be used to generate the simulated data, we used the rbinom function to sample from the appropriate bernoulli distribution.

\subsection{Standard Errors}
Standard errors are similar to the general case however we replace $L^{(1)}(\bm(\theta))$ with the marginal likelihood function $L^{(2)}(\bm(\theta))$ resulting in the following hessian matrix:
\begin{equation*}
    \frac{1}{M}\Sigma_{m=1}^{M}\frac{D^2log(L^{(2)}(\bm{\theta}))}{D^2\bm{\theta}} + 
    \frac{1}{M}\Sigma_{m=1}^{M}(\frac{Dlog(L^{(2)}(\bm{\theta}))}{D\bm{\theta}})^2 -
    [\frac{1}{M}\Sigma_{m=1}^{M}\frac{Dlog(L^{(2)}(\bm{\theta}))}{D\bm{\theta}}]^2
\end{equation*}
\newpage

\section{Data Analysis using 2019 CSL data}
\subsection{Analysis Using the Gamma Frailty Model}
Peng, Hu, Swartz (2024) identified five covariates they believed may have an effect on the time to corner kick, and the probability that a Type 2 kick is Type 2-L. These five variables are definded as follows:

\begin{itemize}
    \item $Z_1 = 1$ if the kick occurred in the first half, and $Z_1 = 0$ if the kick occurred in the second half.
    \item $Z_2 = 1$ if the corner kick is awarded to the home team, $Z_2 = 0$ if it is awarded to the away team.
    \item $Z_3$ is the score differential at the time of the kick, a positive value means the team is leading.
    \item $Z_4$ is the red card differential at the time the kick is awarded, a positive value means the team has more players on the pitch.
    \item $Z_5$ is the decimal odds (European odds) of the team that has been awarded the corner kick. 
\end{itemize}
We began by fitting the model including all five of these covariates in the proportional hazard function for Type 1 and 2-L corner kicks as well as in the logistic regression model for Type 2-L probability.\\

Using the frailty model we found $Z_2$ (indicator of the team being at home) to have a significant positive effect on the hazard function. This aligns with common soccer knowledge that home teams tend to be awarded more corner kicks. We also found that $Z_3$ (score differential) has a significant negative effect on the hazard, which also aligns with common soccer knowledge. When a team is leading they are more concerned with maintaining their lead than they are scoring goals. In order to maintain better defensive structure and hold on to their lead teams will play more conservatively.\\

While Peng, Hu and Swartz (2024) found none of the covariate effects to significantly impact Type 2-L probability. We found three of the five covariates to have significant effects on Type 2-L probability. One possible explanation of this is that using the frailty model results in increased power in determining whether a covariate's effect on Type 2-L probability is significantly different from 0.\\

Table 1 displayed below contains the estimated parameters using the 2019 CSL data.

% In LLNCS, table captions must be placed ABOVE the table. 
\begin{table}[H]
    \caption{Bolded values indicate significance}
    \label{tab:my_label}
    \centering
    \begin{tabular}{|c|c c|}
    \hline
      Parameter &Estimate &Standard Error \\
    \hline
       In model for Type 1 and 2-L corner kicks: \\
       $\theta_w$ (variance of frailty RV) & \textbf{0.247} & \textbf{0.051}\\ 
       $\lambda_1$ (rate) & \textbf{0.021} & \textbf{0.002}\\
       $\gamma_1$ (shape) & \textbf{0.924} & \textbf{0.024}\\
       $\beta_1$ (coef to $Z_1$) & -0.024 & 0.066\\
       $\beta_2$ (coef to $Z_2$) & \textbf{0.172} & \textbf{0.082}\\
       $\beta_3$ (coef to $Z_3$) & \textbf{-0.096} & \textbf{0.029}\\
       $\beta_4$ (coef to $Z_4$) & -0.134 & 0.113\\
       $\beta_5$ (coef to $Z_5$) & -0.019 & 0.015\\
    \hline
       In model for Type 2-S corner kicks:\\
       $\lambda_2$ (rate) & \textbf{1.463} & \textbf{0.062}\\
       $\gamma_2$ (shape) & \textbf{3.542} & \textbf{0.379}\\
    \hline
       Logistic Regression Parameters:\\
       $\alpha_0$ (intercept)& \textbf{1.638} & \textbf{0.086}\\
       $\alpha_1$ (coef to $Z_1$)& 0.122 & 0.228\\
       $\alpha_2$ (coef to $Z_2$)& \textbf{0.296} & \textbf{0.118}\\
       $\alpha_3$ (coef to $Z_3$)& 0.050 & 0.089\\
       $\alpha_4$ (coef to $Z_4$)& \textbf{-0.385} & \textbf{0.069}\\
       $\alpha_5$ (coef to $Z_5$)& \textbf{0.076} & \textbf{0.030}\\
    \hline
    \end{tabular}
\end{table}

The indicator of home team, red card differential, and European betting odds were found to have significant effects on the Type 2-L probability of Type 2 corner kicks. This means under the new model three effects were found to be significant compared to none under the assumption that all corner kicks are independent. This potentially indicates that the frailty model is more powerful than the model in Peng, Hu, Swartz (2024). $94.4\%$ of the simulated $\eta$s from the final iteration of the MCEM algorithm were 1 (indicating a long gap time). This means that just over 1 in 20 corner kicks directly generates another corner.\\ 

We model all Type 2-S corner kicks as weibull random variables. When all covariates are 0 and the frailty random variable is exactly 1 Type 1 and 2-L corner kicks are weibull distributed. However, these corner kicks have drastically different event times. This is exemplified by the drastic difference in means. The Type 1 and 2-L weibull distribution under these strict conditions has a mean time to corner kick of 49.41 minutes. This is contrasted by the very short mean of the Type 2-S corner kicks that is only 36.93 seconds. Such a distinct difference suggests the necessity of a mixture distribution. We visualize this in Figure 1 using the estimated survival functions under these circumstances, which depict two very different survival functions.\\

The bold lines in the plots shown in Figure 2 represent the home and away survival functions when the frailty random variable is exactly 1. Each plot contains an additional 50 faintly drawn lines which depict other survival functions at the same covariate levels but the frailty random variable is sampled from the $Gamma(\frac{1}{\theta_w},\frac{1}{\theta_w})$ distribution. This demonstrates the heterogeneity of corner kicks that are not taken by the same team within the same game. You'll notice that the survival functions of the home team tend to be slightly lower than the away team's at the same time. This aligns with the common soccer knowledge that the home team tends to get more corners.\\

Plot three compares the survival functions estimated using the previous model with the survival functions estimated using the new model.
\subsection{Model Comparison}
To determine if the frailty model ia a significantly better fit than the model that assumes independence of all corner kicks we conduct a likelihood ratio test using the test statistic $\Lambda = -2[log(L^{(old)}(\theta))-log(L^{(2)}(\theta))]$.
The frailty model contains one additional parameter $\theta_w$, so we assume $\Lambda$ has $\chi^2$ distribution with one degree of freedom. Obtaining the estimates using M=200 for the final iteration: $\Tilde{Q}^{(old)}(\bm{\theta}_{old},\bm{\hat{\theta}}_{old}) = -5407.3788$, and $\Tilde{Q}^{(2)}(\bm{\theta},\bm{\hat{\theta}}) = -5386.6376$ giving $\hat{\Lambda} = 41.4823$ that yields a p-value < 0.0001.\\

The Bayesian information criterion (BIC) for the model of Peng, Hu, and Swartz (2024) is 10869.78, while the BIC of the frailty model is 10828.30. A lower BIC is typically associated with striking a better balance between simplicity and model fit. 

\section{Discussion}
In this paper we introduce a frailty model to address the correlation of a team's times to corner kicks within the same game. We then demonstrated that the new model is a significantly better fit to the 2019 CSL data than if independence among all corner kicks is assumed.\\

This papers considers only the case that the corner kicks of the home and away team within the same game are independent of each other, it may be worth investigating corner kicks using a competing risks model to address this. Additionally it may be worth employing a multilevel model to address the potential correlation of a team's corner kicks across different games.\\

Since 2000, sports betting has grown massively partly due to the accessibility of online gambling (Garcia, Perez, Suarez, 2025). The statistical analysis of soccer data allows the increasing number of sports bettors to make informed decisions. Extensive study of the goal scoring process in soccer using bivariate poisson processes already exists, as seen in Maher (1982), Dixon and Coles (1997), and Karlis and Ntzoufras (2003). These models can be applied to inform betting strategy on match results and total goals scored, both of which are popular wagers. The rise of sports betting has also given rise to other boutique wagers, such as corner betting. Notable corner bet is an over/under bet on the number of corner kicks in a game or the over/under on the time elapsed until the first corner kick occurs. A natural extension of this paper is to use the outlined model to inform betting strategy and conduct market efficiency analysis similar to Palssen and Laurens (2023).

\nocite{*}

% LLNCS uses standard BibTeX with splncs04 style [cite: 211, 257]
% \printbibliography <-- Removed
\bibliographystyle{splncs04}
\bibliography{bibliography}

~\\
% \newpage % Commented out for flow
\newpage
\section*{Figures}
The two survival functions plotted together when all covariate levels are fixed at zero.
\begin{figure}
    \centering
    \includegraphics[scale = 0.6]{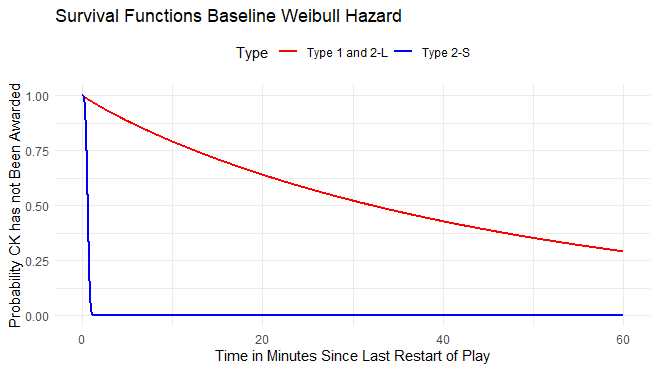}
    \caption{}
    \label{figure 1}
\end{figure}
\newpage
Bolded line is the Type 1 and 2-L survival functions when the frailty variable is exactly 1.
\begin{figure}
    \centering
    \includegraphics[scale = 0.35]{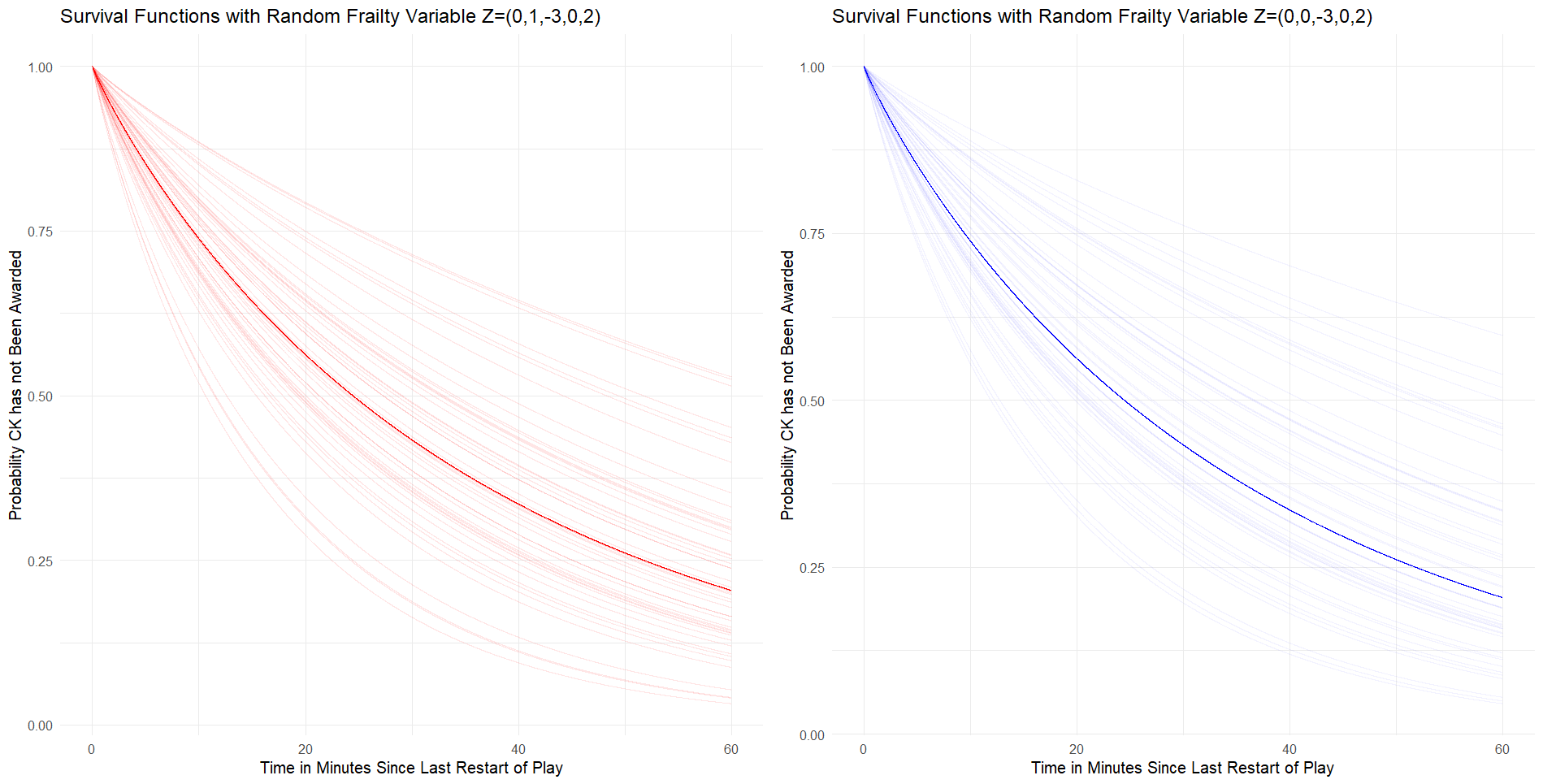}
    \caption{}
    \label{figure 2}
\end{figure}
\newpage
The survival function for different frailty values when covariates are all fixed at zero.
\begin{figure}
    \centering
    \includegraphics[scale = 0.6]{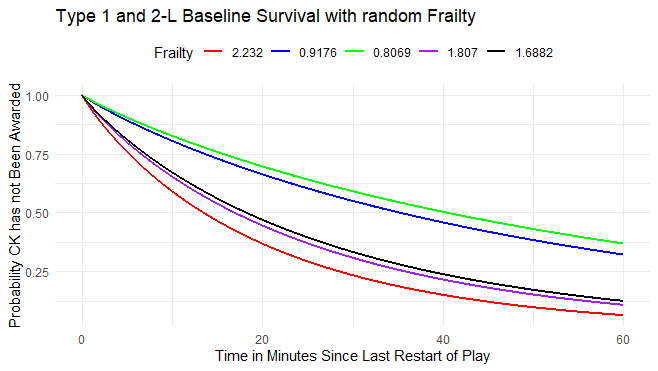}
    \caption{}
    \label{figure 3}
\end{figure}
\end{document}